\documentclass[12pt]{article}
\usepackage{epsfig}

\usepackage{amssymb}

\def\url#1{{\ttfamily\def\/{/\discretionary{}{}{}}#1}}

\begin{document}

\title{Time and entropy.}
\author{Vladimir I. Garaimov\\
\it Astronomy Department, University of Maryland,\\ College Park, MD 20742, USA \\
E-mail: gvi@astro.umd.edu}
\date{}
\maketitle

\centerline{\bf abstract}
This paper presents alternative  ideas on the physics of time that lead to
 a new interpretation of cosmological redshifts.
These ideas are based on the close relationship between the speed of time and 
entropy processes in our universe.
I give numerical estimates and describe laboratory experiments and observational 
effects that can test the new theory.        

\vskip 2cm

\centerline{\bf Postulates}

We will define entropy as the irreversible dispersal of energy and will
regard time as a physical process.

We postulate that:
\begin{itemize}
\item[1] Time has close relationship with entropy. An increase in entropy will cause
a corresponding increase in the speed of time. We can say that the potential 
energy of matter has been converted into the "kinetic energy"(speed) of time.
\item[2] Any object carries information (as an energy level) about the time when it 
was created. 
\end{itemize}

The consequences of these postulates are : 

1. As the entropy of the Universe increases time is accelerated.

Let's analyze a "stationary" entropy process.
During a period of time $dT$ an entropy increase by a factor $m$  results in 
an acceleration of time by a factor $n$ $(n>1)$. The next period $\delta T_1$ 
will be $1/n$ times shorter, but during this period 
the entropy will also increase $m$ times, as 
the process is "frozen" into the time. This results in 
time being  accelerated exponentially.

Let's analyze this from the position of a cosmological redshift.
Assume a  photon is emitted with frequency $hv$, and 
after a certain period of time is registered by a radiation detector.
 During this period the speed of time is accelerated $n$ times. 
Therefore the radiation detector registers the photon with a frequency $hv/n$, 
i.e., the photon is observed with a redshift. It will depend only how long 
the photon has existed, i.e. the longer the time, the greater redshift effect.
In other words, the further from the observer the  source of 
emission is, the greater the redshift, and the principle of 
symmetry is true for this effect. This means that two objects located some 
distance from each other both observe one another to have the same redshift.
Therefore, cosmological redshift may be explained by the effect of time 
acceleration.

\begin{figure}[h]
\psfig{file=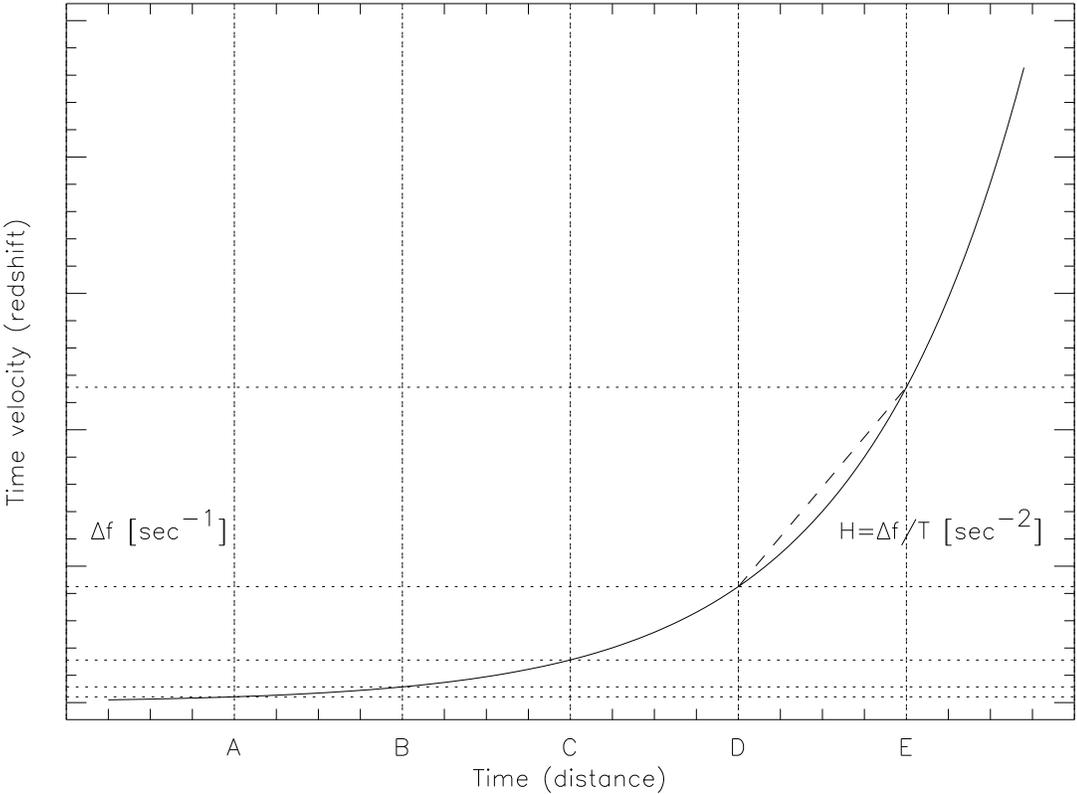,width=1.\textwidth}
\caption{Redshift (time velocity) evolution.}
\end{figure}
2. The Hubble constant is the numerical value of the derivative of the speed of time, 
i.e. it is the acceleration of time (see Fig.1). It has units of
acceleration, i.e. a variation of frequency of radiation $(sec^{-1})$ 
 divided by a time period $(sec)$.
\vskip 0.5cm
If the universe experiences episodes during which entropy jumps rapidly, 
such as the first epoch of star formation, we might expect to see a record of this in 
the variation of the Hubble constant with distance (time).

\vskip 1cm
\centerline{\bf Numerical estimates.}

Let us estimate the time acceleration from the value of the Hubble constant. 
 $$ H\approx 60~km~sec^{-1}~MPc^{-1}$$
The time it takes light to travel 
$1MPc \approx 10^{14} sec$
$$v/c=aT => 60/3\times 10^5= 10^{14}a$$
where: $a$ - acceleration of time; $T$ - time period.
As result a second decreases by \underline{$2\times 10^{-18}sec$} per second.

We can compare this value with the value calculated from the correction to
the equinox after all known precessions are accounted for.
Figure 2 shows the Equinox corrections from solar and lunar observations  [1].

\begin{figure}[h]
\psfig{file=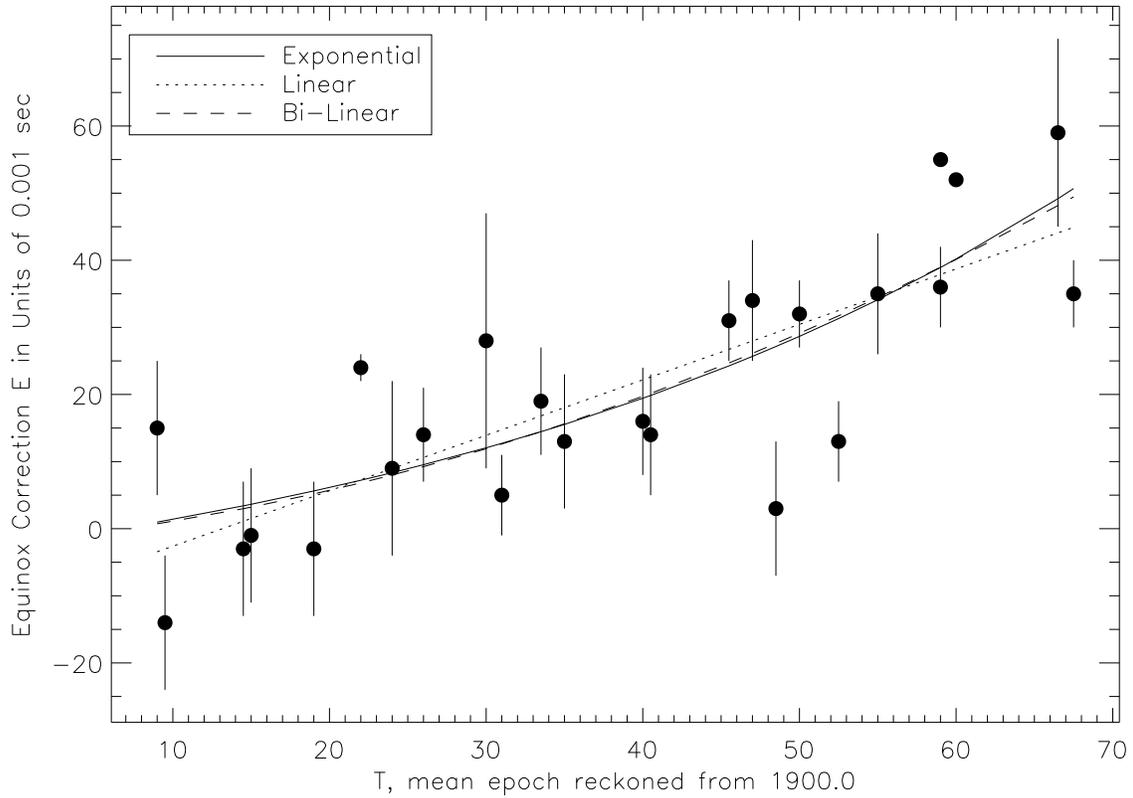,width=1.\textwidth}
\caption{Equinox corrections from solar and lunar observations.
Three different interpolations are also presented, an exponential
interpolation (solid line; sigma =121.87); a linear interpolation 
(dotted line; sigma=127.56);
 bi-linear interpolation (dashed line; sigma=122.42).  
}
\end{figure}
 
The Equinox was determined with the help of chronometers. Our
standards of time are atomic and are "frozen" into time. From
the above discussion it follows that during this time period a second 
decreased, and resulted in the corrections presented in figure 2.

Let's estimate the time acceleration using a linear approximation and the 
formula: $dT=at^2/2 => a=2dT/t^2$.
The numerical data from [1] ($\dot E=+0.00085$ sec per year) 
give \underline{$a\approx 1.7\times 10^{-18}sec$} per second.

Comparing this result with that calculated from the Hubble constant we see
the orders of the values are the same.
\vskip 1cm

\centerline{\bf Observations and experiments.}

\subsection*{Observations}

From Fig.1 we see that:
\begin{itemize}
\item 
Two equal intervals will show different measured redshifts because
it increases non-linearly with the passage of time, i.e for an  
observation of a single object the redshift will increase non-linearly with time
( see Supernovae project result: \url{http://www-supernova.lbl.gov/});

\item Hubble constants calculated for objects located at different 
distances are not equal (the further away an object is the smaller the Hubble 
constant, i.e.$ H_{DE}>H_{CE}>H_{BE}>H_{AE}$).
\end{itemize}

\subsection*{Laboratory experiments.}

A localization of entropy processes suggest that the increasing
 speed of time (its acceleration) is also localized in space and 
decreases with distance. On the basis of these ideas the following 
experiment has been devised (see.Fig 3).

The experimental technique is as follows:
\begin{itemize}
\item[] a) measure the frequency of radiation from the laser (2)
\item[] b) heat the liquid in the vessel (1)
\item[] c) measure the frequency of radiation from the laser (2) while 
the liquid is boiling
\end{itemize}

\begin{figure}[h]
\psfig{file=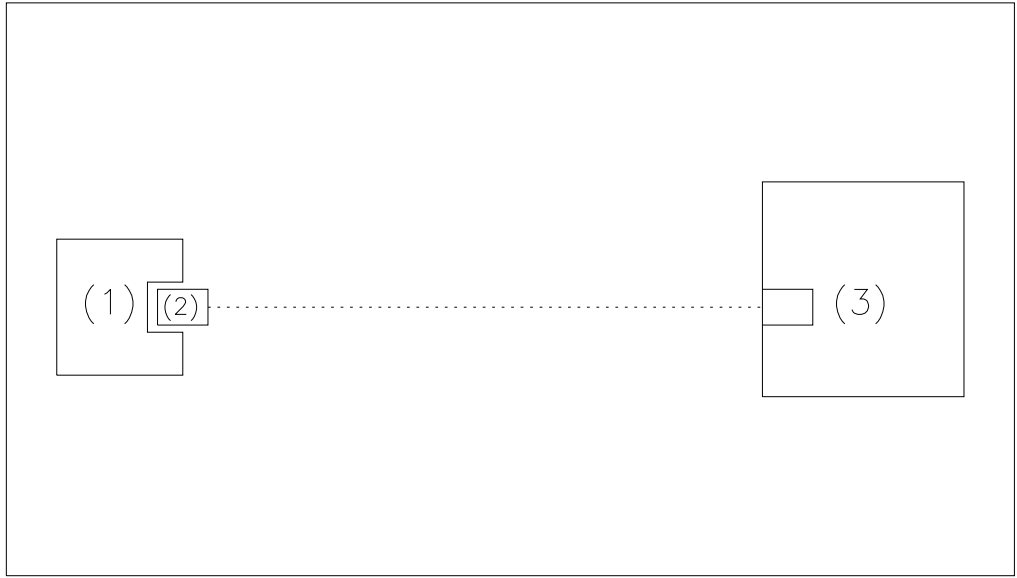,width=1.\textwidth}
\caption{An experimental device for measuring entropy blueshift.
1 -- entropy source (vessel containing a boiling liquid); 
2 -- laser source(milliwatt); 
3 -- high resolution spectrometer.}
\end{figure}
The idea of the experiment consists of the following:

the boiling liquid in the vessel increases the entropy locally, i.e.
it accelerates time. The radiation from the laser source is generated at
the faster time while the spectrometer measures the radiation at the slower 
time.
As a result the radiation registered from laser source during the boiling of 
the liquid has a higher frequency than without boiling the liquid.
Increasing the intensity of the boil (generating more entropy) will cause
the frequency of observed radiation to increase.
For this technique a blueshift of the laser radiation will be observed.
If the entropy system is located close to the spectrometer a redshift of 
the laser radiation will be observed.

Also this effect should be observable using spectral absorbtion lines.
\vskip 0.3cm

The same experiment we can do into the interplanetary space.
The Sun is a huge entropy system. The satellite transmitter is a 
coherent radiation system. Within the bounds of these ideas, unusual
redshifts from the radio beacons on the distant spacecraft ("Pioneer" 
and "Voyager") will be observed.  
 
\vskip 0.5cm

Also we can observe the cumulative effect of time acceleration.
For example, particle collisions (inside a particles accelerator) or 
nuclear explosions are  very powerful entropy processes. If one of two 
synchronized chronometers is located close to this system and another is
located very far away, then after the collision or the explosion the closest
chronometer will register more time than the other. It will be because during the
collision or explosion the first chronometer will be into the fast speed time zone.   

\vskip 1cm
\centerline{\bf Conclusion}
The ideas presented here are speculative. However, this theory can be
applied on any spatial scale, e.g.: extra-galactic scales
(cosmological redshift), planetary scales (equinox correction) and
 local scales (laboratory experiments).
The results of the laboratory experiment will provide a genuine test
of the correctness of this theory.

\end{document}